\documentstyle[12pt]{ioplppt}  

\begin{document}
\jl{1} 
\title{
About the Dirac Equation with a $\delta $ potential}
\author{R. D. Benguria, H. Castillo\footnote{On leave of absence from 
\em { Pontificia Universidad Cat\'olica del Per\'u,} 
 \rm Lima, Per\'u.}, M. Loewe}
\address{
Facultad de F\'{\i}sica, Pontificia Universidad Cat\'olica
de Chile, Casilla 306, Santiago 22, Chile}

\begin{abstract}
An elementary treatment of the Dirac Equation in the presence
of a three-dimensional spherically symmetric $\delta (r-r_{0})$-potential
is presented. We show how to handle the matching conditions
in the configuration space, and discuss the occurrence of
supercritical effects.
\end{abstract}
\maketitle

The problem of solving the Dirac equation in the presence of a
$\delta $ potential represents a curious and non simple situation, in 
contrast with the equivalent problem in non-relativistic quantum mechanics,
i.e., the Schr\"{o}dinger equation,
which is discussed in any course on Quantum Mechanics.
This is basically related to the fact that being the Dirac equation of
first order, a singular potential, like the $\delta $ one,
induces discontinuities at the level of the wave function themselves 
instead of the usual discontinuities that appear
in the first derivative in the Schr\"{o}dinger scenario.

This puzzling situation has been discussed previously in the literature
by many authors. Rigorous construction 
of self adjoint extensions for the Dirac operator,
allowing the handling of matching conditions at the support 
of the $\delta $ potential were discussed in \cite{ditt}. Using their
results, in \cite{magd} a discussion of this problem was presented.
However, we have realized that the proposed solution corresponds to
a different self adjoint extension, associated to also a different
singular potential.

At present, a simple and elementary discussion,
without the necessity of invoking sophisticated
mathematical constructions,
is still not available in the literature. In this note
we want to overcome this situation, showing
in elementary terms how to handle the problem in the configuration space.
Other authors have shown how to handle the problem in momentum space
\cite{danilo}.
Using our results, we  discuss the occurrence of supercritical effects,
i.e., the possibility that  the ground state starts to dive 
into the depths of the Dirac sea, implying positron emission \cite{gr0}.

The Dirac  equation with an external potential can be written as
\begin{equation}
H \psi = E \psi,
\end{equation}
\noindent where 
\begin{equation}
H = c \hat \alpha \cdot \hat p + \hat \beta m c^2 + V(r) .
\end{equation}
Here $\hat \alpha $ and $\hat \beta$ are the usual $ 4 \times 4$ Dirac matrices 
and $\psi$ is the four-component Dirac spinor.

In what follows we will focus on the attractive spherically 
symmetric (vector) potential given by

\begin{equation}
V(r) = -a \delta (r-r_{0}),
\end{equation}
\noindent with $a >0$.

It is convenient to recall some general properties of the solution
of the Dirac equation in a central potential. 
For more details the reader may consult
the book by Greiner, M\"uller, and Rafelski \cite{grein} and reference
therein. In this case,
the complete set of commuting operators is given by 
$H$, $J^{2}$, $J_{3}$ and $K $, where $\vec J$ is the total angular
momentum (i.e., $\vec J = \vec L + \vec S $), and $\vec K$  is defined by

\begin{equation}
K= \beta ( \vec{\Sigma } \cdot \vec{L} + \hbar \hbox
{1\negthinspace \negthinspace 1}).
\end{equation}

In this expression $\vec{\Sigma }=\left( \begin{array}{cc} \vec \sigma & 0 \\
0 & \vec \sigma \end{array} \right)$ where the $\sigma$'s are the Pauli
matrices. In terms of $\vec \Sigma$, $\vec S = \frac \hbar 2 \vec \Sigma$. 
On the other hand $\vec L $ is the orbital angular momentum. The eigenvalues of the
operator $K $ are given by

\begin{equation}
K\psi = -\kappa \hbar \psi = \pm (j + \frac{1}{2})\hbar \psi,
\label{kpi}
\end{equation}
\noindent where $\psi = \left( \begin{array}{c} \psi_u \\ \psi_l \end{array} 
\right)$ is the four-component spinor solution of the Dirac equation written in terms
of the two-component upper and lower $\psi_u$ and $\psi_l$.   
In the equation \eref{kpi}, $j=0,\frac{1}{2},1,\frac{3}{2},...$ are the usual eigenvalues
of $J^{2}$ according to:
\begin{equation}
J^{2}\psi = j(j+1)\hbar^{2} \psi.
\end{equation}

We note that the four-component Dirac spinor is not an eigenfunction of $L^2$.
However the upper and lower components, taken separately, satisfy
\numparts
\begin{equation}
L^2 \psi_u(x) =[j(j+1)\hbar^2+\kappa \hbar^2+\frac{1}{4}\hbar^2]\psi_u(x)
\equiv l_u(l_u+1)\hbar^2 \psi_u(x),\\
\end{equation}
\noindent
and
\begin{equation}
L^2 \psi_l(x) =[j(j+1)\hbar^2-\kappa \hbar^2+\frac{1}{4}\hbar^2]\psi_l(x)
\equiv l_l(l_l+1)\hbar^2 \psi_l(x).
\end{equation}
\endnumparts
Note that the orbital parities of the upper and lower components have 
opposite signs. It is convenient to parametrise the four-component spinor 
by separating the radial and angular dependence according to 

\begin{equation}\Psi =
\left[ \begin{array}{c}\psi _u \\ \psi _l\end{array}\right] =
\left[ \begin{array}{c} g(r) \Omega_{j l_u m}(\theta,\phi)\\
                        i f(r) \Omega_{j l_l m}(\theta,\phi)\end{array}
\right],\end{equation}

\noindent where 
\begin{equation}
\Omega_{jlm}(\theta,\phi)= \sum_{m',m_s} \left(l s j|m' m_s m \right)
 Y_{lm'}(\theta,\phi) \chi _{\frac 1 2 m_s},
\end{equation}
\noindent are the spherical spinors that carry out the angular part. Here we have coupled, through appropriate Clebsh-Gordan coefficients, the scalar spherical harmonics $Y_{lm'}(\theta,\phi) $ with the eigenfunctions of the spin given by the two-components spinors $\chi_{\frac 1 2 m_s}$.

For the present discussion we do not need the explicit form of the
angular part. If we now consider the Dirac equation for a spinor 
parametrised in this way, it is not difficult to show that the radial
components satisfy the following set of coupled differential equations:
\numparts
\begin{equation}
\hbar c \left( \frac{\d F}{\d r}-\kappa \frac F r \right) =
-\left(E-V(r)-mc^2\right) G(r),                        \label{edo1}\\
\end{equation}
\noindent
and
\begin{equation}
\hbar c \left( \frac{\d G}{\d r}+\kappa \frac G r \right) =
 \left(E-V(r)+mc^2\right) F(r),                        \label{edo2}
\end{equation}
\endnumparts
\noindent
where $G(r)\equiv rg(r) $ and $F(r)\equiv rf(r)$.
Because of the linear discontinuity of the spinor function, given by the
delta potential, we need to fix the boundary conditions in the neighbourhood
of the shell $r=r_0$. If we multiply equation \eref{edo1} by $F$, 
equation \eref{edo2} by
$G$, and then, sum both expressions to remove the singular delta function,
we obtain:
\begin{equation}
F^{\prime} F+G^{\prime}G=\frac{2mcFG}\hbar +\kappa \frac{\left( F^2-G^2\right) }{%
r\hbar c}
\end{equation}

In the previous expression,
the primes denote radial derivatives. By integrating between 
$r_0-\varepsilon $ and $r_0+\varepsilon $ and taken then
the limit when $\varepsilon \rightarrow 0$ we get:

\begin{equation}
\lim_{\varepsilon \rightarrow 0}\int_{r_0-\varepsilon }^{r_0+\varepsilon
}\left( F^{\prime}F+G^{\prime}G\right) \d r=\lim_{\varepsilon \rightarrow
0}\int_{r_0-\varepsilon }^{r_0+\varepsilon }\left( \frac{2mcFG}\hbar +\kappa 
\frac{\left( F^2-G^2\right) }{r\hbar c}\right) \d r
\end{equation}
Assuming that the  discontinuities of these functions are well behaved
we find that: 

\begin{equation}
\lim_{\varepsilon \rightarrow 0}\left( F^2+G^2\right) \bigm| _{r_0-\varepsilon
}^{r_0+\varepsilon }=0.\label{norma}
\end{equation}

We may consider $F$ and $G$ as the real and imaginary parts of a function in $\rm C \negthinspace \negthinspace \negthinspace  \negthinspace 1$  . 
In this context, equation \eref{norma} expresses the fact that the absolute value 
of this function is constant when crossing the support of the $\delta$-potential.
This is in agreement with the condition established by J. Dittrich, P. Exner, and 
P. \u{S}eba (see equation (3.5a) in \cite{ditt}; see also the remarks after equation (21)
in \cite{magd}). In fact, the absolute value of this function is continuous for all $r$.

If we set $F_{+,-}\equiv F(r_0\pm \varepsilon )$ and 
$G_{+,-}\equiv G(r_0\pm \varepsilon )$, \eref{norma} becomes 

\begin{equation}
{F_{+}}^2+{G_{+}}^2={F_{-}}^2+{G_{-}}^2.
\end{equation}

Now, as a second step, we multiply the differential equations by $G$ and $F$,
respectively, and substract them to get 
\begin{eqnarray}
\fl F^{\prime}G-F G^{\prime}=
-\frac{\left( E-mc^2\right) }{\hbar c}G^2+\frac{\left(
E+mc^2\right) }{\hbar c}F^2 
+2\frac {\kappa GF} {\hbar cr}-\frac a{\hbar
c}\delta \left( r-r_0\right) \left( F^2+G^2\right).
\end{eqnarray}

Dividing by $F^2+G^2,$ which is continuous for all values of $r$, we can integrate
in the neighbourhood of the shell radius

\begin{equation}
\lim_{\varepsilon \rightarrow 0}\int_{r_0-\varepsilon }^{r_0+\varepsilon }%
\frac{ F^{\prime}G-F G^{\prime}}{\left( F^2+G^2\right) } \d r=-\frac a{\hbar
c}\lim_{\varepsilon \rightarrow 0}\int_{r_0-\varepsilon }^{r_0+\varepsilon
}\delta \left( r-r_0\right) \d r.
\end{equation}

By using 
\begin{equation}
\frac{F^{\prime}G-F G^{\prime}}{\left( F^2+G^2\right) }=\frac 1{\left( F/G\right) ^2+1}
\frac \d {\d r} \left( \frac F G \right),
\end{equation}
and since $\int {\frac 1{1+h^2(x)}\d [h(x)]}=\arctan (h(x))$ we have:

\begin{equation}
\lim_{\varepsilon \rightarrow 0}\left( \arctan \frac{F(r)}{G(r)}
\right) \vert_{r_0-\varepsilon }^{r_0+\varepsilon }=-\frac a{\hbar c}.
\end{equation}

In this way, our second boundary condition can be written as

\begin{equation}
\arctan \frac{F_{+}}{G_{+}}-\arctan \frac{F_{-}}{G_{-}}=-\frac a{\hbar c}.
\end{equation}

Expressing the coupling constant in units of $\hbar c$, we define the dimensionless
parameter
$\alpha \equiv \tan( a /{\hbar c})$. Our second boundary condition
can then be expressed as

\begin{equation}
\frac{F_{+}}{G_{+}}=\frac{({F_{-}}/{G_{-}})-\alpha }{1+ \alpha 
({F_{-}}/{G_{-}})} \label{bc}
\end{equation}

Except for an arbitrary phase, the last expression can
be written as a matricial relation between the radial functions
at both sides of the potential,

\begin{equation}
\left[\begin{array}{cc} F_{+} \\ G_{+} \end{array} \right]
=\left[ 
\begin{array}{cc}
\cos (a/\hbar c)  & -\sin (a/\hbar c) \\ 
\sin (a/\hbar c)  & \cos (a/\hbar c)
\end{array}
\right]
\left[\begin{array}{cc} F_{-} \\ G_{-} \end{array} \right]
\equiv A \left[\begin{array}{cc} F_{-} \\ G_{-} \end{array} \right]
\end{equation}

This matrix $A$ is unitary (actually orthogonal), $\det A=1$, and
contains the information for finding
the eigenvalue equation for the bound states. 
Returning to our complex valued function with real and imaginary part 
given by $F$ and $G$, respectively, it
is interesting to remark that the $\delta $ manifests
itself through a change of phase of this function,
given by $\tan (a/\hbar c)$. 

For the solutions of equations \eref{edo1} and \eref{edo2} corresponding to the free case, 
we may separate the space into two regions: 

Region I, $r<r_0$ ,

\begin{eqnarray}
G_{I}(r) &=&A_I r \left( \frac \pi {2kr}\right)  
^{1/2} {I}_{l_\kappa +1/2} (kr), \\
F_{I}(r)&=&A_I\frac{k\hbar c}{E+mc^2}r
\left( \frac \pi{2kr}\right) ^{1/2} {I}_{l_{-\kappa} +1/2}(kr).
\end{eqnarray}
Region II, $r>r_0$,
\begin{eqnarray}
G_{II}(r) &=&A_{II}r\left( \frac \pi  {2kr}\right) ^{1/2}
{K }_{l_\kappa +1/2}(kr), \\
F_{II}(r) &=&-A_{II}\frac{k\hbar c}{E+mc^2}r\left(\frac \pi{2kr}\right)  
^{1/2} {K}_{l_{-\kappa }+1/2}(kr).
\end{eqnarray}
The relations we are looking for reduce to

\begin{eqnarray}
\frac {F_{I}} {G_{I}} &=& \frac {k \hbar c}{E+m c^2} 
\frac {{I}_{l_{-\kappa}+1/2} (kr)} {{I}_{l_{\kappa}+1/2} (kr)} \label{f1}\\
\frac {F_{II}} {G_{II}} &=& -\frac {k \hbar c}{E+m c^2} 
\frac {{K}_{l_{-\kappa}+1/2} (kr)} {{K}_{l_{\kappa}+1/2} (kr)} \label{f2}
\end{eqnarray}

Taking into account that for the ground state, for $j=l+s=1/2$,
we have ${l}_{\kappa}=0$ and ${l}_{-\kappa}=1$ we can write

\begin{eqnarray}
I_{1/2}(kr)&=&\sqrt{\frac 2{\pi kr}}\sinh (kr) \\
I_{3/2}(kr)&=&\sqrt{\frac 2{\pi kr}}
\left( \cosh \left( kr\right) -\frac{\sinh (kr)}{kr}\right) \\
K_{1/2}(kr)&=&\sqrt{\frac \pi {2kr}}\e^{-kr} \\
K_{3/2}(kr)&=&\sqrt{\frac \pi {2kr}}\e^{-kr}
\left( 1+\frac 1{kr}\right)
\end{eqnarray}

In these equations  $k $ denotes a wave number, 
$\hbar ck=\sqrt
{m^2c^4-E^2}$. In order to find the eigenvalues of the hamiltonian, we evaluate \eref{f1} and \eref{f2} of $r_0$ and use the boundary condition \eref{bc}. In this way we are led to solving the following transcendental equation
 
\begin{eqnarray}
\fl \nonumber \frac {k \hbar c}{E+m c^2} \left( 1+ \frac 1{kr_0} \right) + \alpha
\left({\frac {k \hbar c}{E+m c^2}}\right)^2 \left( 1+ \frac 1{kr_0} \right)
\left( \frac {1 - \tanh (kr_0)} {\tanh (kr_0)} \right) \\
\lo= \frac {k \hbar c}{E+m c^2} \left( \frac {1 - \tanh (kr_0)} {\tanh (kr_0)}
 \right)- \alpha \label{autoval}
\end{eqnarray}

Without solving explicitly this equation, we can analyse the behaviour of the ground state  energy  $E$, as a function of the parameter $\alpha = \tan (a / \hbar c)$, related to the coupling constant $a$. We start by introducing the following dimensionless variables
\numparts
\begin{eqnarray}
\varepsilon &\equiv& \frac E {mc^2}, \\
\rho &\equiv& \frac {r_0} {\hbar / mc}, \\
s_0&\equiv&\rho \sqrt {1-\varepsilon^2}, \\
u_0&\equiv&\rho (1+\varepsilon),\\ \label{ff}
g_0&\equiv&\frac {\tanh (kr_0)} {kr_0}. \label{g0}
\end{eqnarray}
\endnumparts
In terms of these variables, our eigenvalue equation \eref{autoval} can be written as

\begin{equation}
\alpha = \frac {s_0 u_0 (1+g_0 s_0) } {{u_0}^2-(s_0+1)s_0(1-g_0)}. \label{avn}
\end{equation}

For a fixed value of the radius of the $\delta$ -shell, $r_0$ (with $r_0 \neq 0$), we are interested in determining the existence of the ground state  energy in the interval 
$(-mc^2,mc^2)$ (i.e., $-1 \leq \varepsilon \leq 1$). The existence of an $\varepsilon$ in this range will depend on the values of the coupling constant $a$ through the dimensionless parameter $\alpha$.

Since $\hbar c k = \sqrt{m^2 c^4- E^2} = mc^2 \sqrt{1- \varepsilon ^2}$, we get from \eref{g0} that
\begin{equation}
\lim _{ \varepsilon \rightarrow \pm 1} g_0=1. \label{s5}
\end{equation}
On the other hand, using \eref{ff} we get 
\begin{equation}
\lim _{ \varepsilon \rightarrow -1} \frac {1 -g_0} {u_0} = \frac 2 3 \rho. \label{s6}
\end {equation}

Using the limits \eref{s5} and \eref{s6} in Eqn. \eref{avn} we see that as $E$ approaches the free state (i.e., $\varepsilon \rightarrow 1^{-}$), $\alpha \rightarrow 0^{+}$, which agrees with the fact that the bound states disappears for a vanishing potential.

Proceeding as before, with  $\varepsilon \rightarrow -1^{+}$ (i.e., as the energy approaches the Dirac sea), we infer from equation \eref{avn} that $\alpha$ approaches the value $-3/{2 \rho}$, which gives 
\begin{equation}
\tan \left( \frac {a_{crit}} {\hbar c} \right) = -\frac 3 {2 \rho}, \label{crit}
\end{equation}
\noindent where $a_{crit}$ , the minimum positive solution of \eref{crit}, is the value of the coupling constant for which the ground state energy sinks into the Dirac sea (e.g., for $\rho = 1$, $a_{crit}=2.19 \hbar c$). 

The numerical solution of equation \eref{avn} for the dimensionless ground state energy $\varepsilon$ as a function of $\rho$ and $a$ is plotted in Figure 1. Notice that for a fixed value of $\rho$, $\varepsilon$ is a decreasing function of $a$. For all finite values of $\rho$ there are supercritical effects. Clearly, the value of $a_{crit}$ for which $\varepsilon$ sinks into the Dirac sea, is increasing with $\rho$. However, the limit $r_0 \rightarrow 0 $ is not well defined in \eref{avn} and thus we cannot find solutions for bound sates in this limit. We would like to remark that a general theorem by Svendsen, \cite{svend,nogami},
tells us that supercritical effects are absent in this limit.

\ack {R. B. acknowledges support from Fondecyt, under grant 1990427.
M. L. acknowledges support from Fondecyt, under grant 1980577. H. C. acknowledges
financial support from a PUC fellowship.}

\pagebreak

\Figure {
In Fig.1, we show the behaviour of the ground state energy
 $ \varepsilon = E/mc^{2}$,
for different values of the size $\rho = r_{0}/(\frac{\hbar}{mc})$  of the 
delta-shell as a function of the dimensionless coupling constant 
$ A =a/(\hbar c)$: $\rho = 0.5$ (dot line),
$\rho = 1.0$ (solid line), $\rho = 2.0$ (dashed line), $\rho = 10.0$ (dot-dashed line).
}

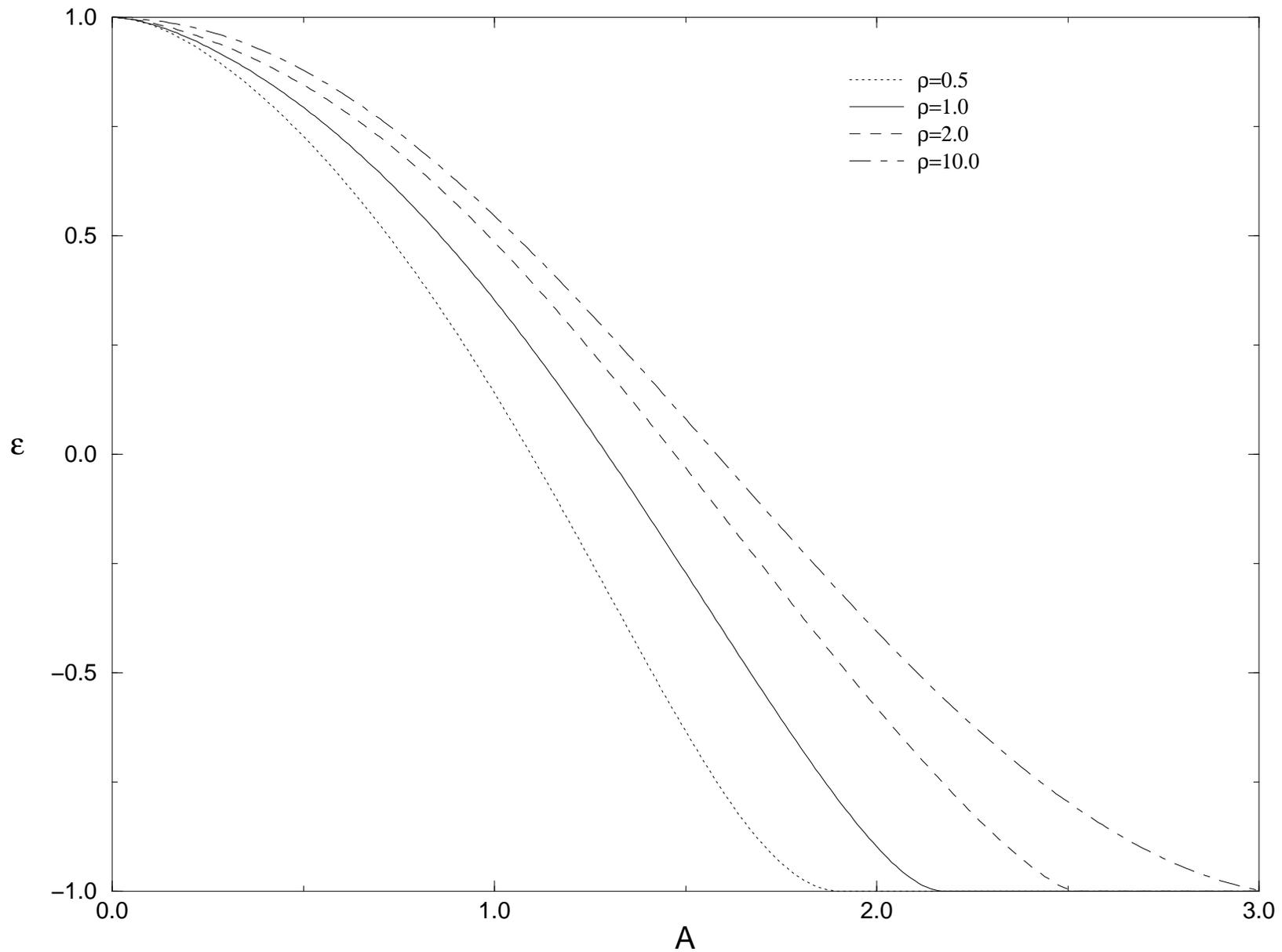
\end{document}